# Effect of columnar defects on the pinning properties of NdFeAsO$_{0.85}$ conglomerate particles

J.D. Moore, L.F. Cohen, Y. Yeshurun*, A.D. Caplin, K. Morrison, K.A. Yates
*Blackett Laboratory, Imperial College, London SW7 2AZ*

C.M. McGilvery, J.M. Perkins, D.W. McComb
*Materials Department, Imperial College, London SW7 2AZ*

C. Trautmann
*Gesellschaft für Schwerionenforshung, Materials Research Group, Darmstadt, Germany*

Z.A. Ren, J.Yang, W.Lu, X.L. Dong, Z.X. Zhao
*National Laboratory for Superconductivity, Institute of Physics and Beijing National Laboratory for Condensed Matter Physics, Chinese Academy of Sciences, PO Box 603, Beijing 100190, PR China*

* Permanent address: Department of Physics, Bar-Ilan University, Ramat-Gan 52100, Israel.

**Abstract**

Oxypnictide superconductor NdFeAsO$_{0.85}$ sample was irradiated with 2 GeV Ta ions at a fluence of 5×10$^{10}$ ions/cm$^2$. High resolution transmission electron microscopy study revealed that the irradiation produced columnar-like defects. The effect of these defects on the irreversible magnetisation in polycrystalline randomly oriented fragments was studied as a function of field angle and field sweep rate. We find that the critical current density is enhanced at fields below the matching field (~1 Tesla) but only marginally. The pinning enhancement is anisotropic and maximum along the defect direction at high temperatures but the pinning then becomes more isotropic at low temperatures. The creep rate is suppressed at high temperatures and at fields below the matching field, indicating the columnar defects are efficient pinning sites at these $H$ and $T$ conditions.

**1. Introduction**

It is hoped that the newly discovered, iron-based oxypnictide superconductors will show promise for technological applications where the basic requirements are to carry high currents in high magnetic fields and to work at relatively high temperatures where the associated cooling costs are lower. The superconducting transition temperature, $T_c$, in these materials has been recorded up to ~55 K (1) which puts the $T_c$ of the pnictides above that of MgB$_2$ superconductors ($T_c$ = 39 K), but below high temperature anisotropic superconductors (HTSC) such as YBCO ($T_c$ = 92 K) and BSCCO ($T_c$ = 110 K). The NdFeAsO$_{0.85}$ sample studied here is a member of the "1111" phase of oxypnictide superconductors. The 1111 phase has been shown to have superconducting critical current densities, $J_c$, of 10$^5$ - 10$^6$ A/cm$^2$ (at 5 K) in single crystals (2;3), while in a Co-doped BaFe$_2$As$_2$ single crystal, a member of the "122" family (4), a $J_c$ of 10$^6$ A/cm$^2$ (at 2 K) has also been reported. In polycrystals of 1111 the $J_c$ is only up to 10$^4$ - 10$^5$ A/cm$^2$ (at 5 K) (5-7), while initial attempts to make wires of 1111 phase using the powder-in-tube method (8) have also shown low intergrain $J_c \sim 10^5$ A/cm$^2$ (at 5 K). So connectivity might be even more of a concern in these materials.

One way to improve $J_c$ is by deliberately introducing defects to the crystal lattice which can then act as strong pinning sites for vortices, increasing the irreversible magnetization and suppressing magnetic relaxation. Additionally the defects can reduce the coherence length thereby increasing the field at which the irreversible magnetization disappears ($H_{irr}$) and increasing $J_c(B)$.

There have been various studies that have aimed to increase the pinning in both 1111 and 122 pnictide superconductors. The effects of neutron irradiation, which introduces point defects to a sample, have been studied for polycrystalline samples of the 1111 phase (5;9). It was shown that $J_c$ is increased and the low temperature $H_{c2}$ was enhanced in polycrystals of SmFeAsO$_{1-x}$F$_x$ (5). However, the disorder induced by the neutrons also depressed the $T_c$ (9). Very recently, the effect of heavy ion irradiation (using 200 MeV Au ions) has been investigated in single crystal 122 with composition Ba(Fe$_{0.93}$Co$_{0.07}$)$_2$As$_2$ and has demonstrated that columnar defects were introduced and act as effective pinning centres, increasing $J_c(B)$ by almost a factor ten compared to the unirradiated crystal (4). From measurements of magnetic relaxation, Nakajima *et al.* infer that the effective pinning energy is enhanced by a factor two with irradiation which is consistent with the increased $J_c$ and reduced vortex creep rates. The superconducting anisotropy is thought to be considerably different for 1111 and 122 compounds, with 1111 reported to be more anisotropic ($\gamma_{Hc2} \sim 5.5$ close to $T_c$ and temperature dependent), the precise value and its temperature dependence still being discussed in the literature (10;11). The 122 oxypnictide systems on the other hand are less anisotropic ($\gamma_{Hc2} \sim 2$ and only slightly temperature dependent), and therefore directional pinning from correlated defects would not be expected to show any strong temperature dependence in this case (2).

Here we study the influence of columnar defects on the more anisotropic 1111 system, where columnar-like pinning defects might be thought to play a more significant role. We present data on heavy ion irradiation of ~100 μm sized polycrystalline conglomerate particles of NdFeAsO$_{0.85}$. We irradiated these samples with 2 GeV Ta ions and studied the field dependence of the magnetization and magnetic relaxation while investigating the effect of changing the field angle away from the defect direction. In a previous study we showed that unirradiated conglomerate particles of NdFeAsO$_{0.85}$ displayed imperfect intergranular connectivity with $J_c \sim 5 \times 10^4$ A/cm$^2$ at 15K (6). We find a marginal increase in $J_c$ with irradiation, but also a temperature-dependent directionality to the pinning enhancement. We discuss the pinning by columnar defects in NdFeAsO$_{0.85}$ from the view-point of pinning in the high temperature anisotropic superconductors.

## 2. Experimental methods

NdFeAsO$_{0.85}$ was made by high pressure synthesis as described elsewhere (1). The material was prepared as a pressed bar and subsequently broken in to platelet-like fragments with a range of sizes. We selected those fragments with an average diameter of ~100 μm and thickness ~50 μm. Structural characterisation has shown the particles are polycrystalline (the grain size is of order microns) with secondary phases of Fe$_x$O$_y$ and NdO (6). The superconducting phase is compositionally homogenous, but there is a small Neodymium oxide phase present that likely contributes to the paramagnetic background seen in magnetization curves.

A large collection of the fragments was laid flat on an aluminium sheet and stuck down with a thin coating of GE varnish. Irradiation was carried out at the GSI Centre for Heavy Ion Research in Darmstadt using a 2 GeV beam energy of Ta ions at a fluence $5 \times 10^{10}$ ions/cm$^2$, i.e. the matching field is 1 T. The beam was directed perpendicular to the aluminium plate such that the Ta ions would be expected to penetrate the full fragment thickness. The collection of fragments was measured while still on the aluminium plate to maintain the defect orientation. The magnetization versus applied field (*M-H*) measurements were performed in an Oxford Instruments variable-temperature Transverse Vibrating Sample Magnetometer (TVSM) which has a rotation stage to allow the sample to be tilted with respect to the applied field direction. The dynamic creep rate, *S*, was obtained using the method known as 'sweep creep' (12) as first proposed by Pust *et al*. (13). The flux creep was determined at each field by taking full *M-H* loops at several field sweep-rates and using the relation $S = d(\ln \Delta M)/d(\ln H/dt)$. The $T_c$ in the irradiated fragments was determined from magnetization data as being ~ 46 K, which we note is the same as $T_c$ in the unirradiated fragments as also extracted from magnetization measurements.

## 3. Results and discussion

A typical NdFeAsO$_{0.85}$ fragment is shown in the optical micrograph in Fig. 1(a). Inspection with field-emission scanning electron microscopy (FE-SEM) in part (b) of the same figure shows that each fragment is made up of plate-like particles of homogenous superconducting material. After irradiating a collection of these fragments with 2 GeV Ta ions we performed high resolution transmission electron microscopy (TEM) analysis to identify the effect of the irradiation. The top surface is the one that was irradiated and we then prepared cross-sections that were oriented in a plane perpendicular to the surface using a focused ion-beam (FIB). Figure 1(c) and 1(d) show TEM cross-sections taken 1 μm and 5 μm below the surface, respectively. Both images show the presence of collinear tracks which, after considering the thickness of the cross-sections, agree in number with the Ta ion fluence of $5 \times 10^{10}$ ions/cm$^{-2}$. By inspection, the tracks are approximately 5 – 10 nm wide and therefore might be expected to suitably pin vortex lines where the vortex radius is set by the superconducting coherence length measured as $\xi \sim 3.5$ nm in the closely related system LaFeAsO$_{0.89}$F$_{0.11}$ (14).

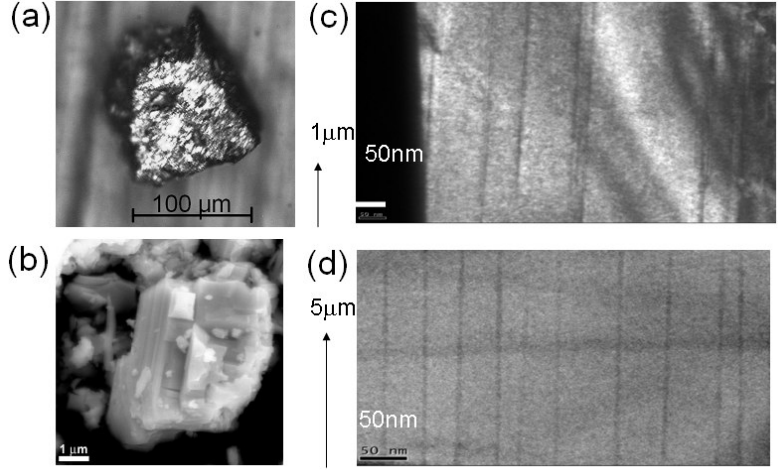

Figure 1. (a) Optical image of a typical NdFeAsO$_{0.85}$ conglomerate particle, first taken from the sintered body and then irradiated by 2 GeV Ta ions with a fluence 5×10$^{10}$ ions/cm$^2$. (b) FE-SEM image of platelet-like NdFeAsO$_{0.85}$ particle. TEM images show (c) a cross-section taken ~1 μm below the irradiated surface and (d) a cross-section taken more than 5 μm below the surface. Both (c) and (d) images show the presence of collinear columnar tracks, which are parallel to the irradiation direction (from top to bottom).

In Fig. 2(a) we show the four-quadrant *M-H* envelope loop at 10 K for the irradiated fragments (red dashed line) and a collection of similarly prepared unirradiated fragments (solid curve). The field was applied parallel to the irradiation direction (we define this orientation as $\theta = 0$ deg). The unirradiated set was additionally measured at different applied field angles and we observed no significant angular dependence. Both curves show the presence of a similar paramagnetic background which was attributed to small amounts of Nd-oxide present in the sample (6). The curve for the irradiated set in Fig. 2(a) shows greatest enhancement of the irreversible magnetisation $\Delta M$ below the matching field of 1 Tesla, which is when the density of vortices equals the density of columnar tracks. We observed a small enhancement in $\Delta M$ at fields up to a few times the matching field of 1 T, which is not unexpected as vortex-vortex interactions mean that the column defects can still affect pinning at fields greater than the matching field (15). We extract $J_c$ from $\Delta M$ using the Bean model and assuming that the fragments are connected and have a cylindrical geometry with average diameter 100 μm. We ignore the $J_c$ in zero field due to asymmetry in the unirradiated *M-H* loop and compare instead the $J_c$ at *H'*=0.25 T for the irradiated and unirradiated sets (*H'* is indicated by the vertical dashed line in Fig. 2(a)). Note that the self-field of the 10 K data shown in Fig. 2(a) is ~0.15T. Figure 2(b) shows the temperature dependence of $J_c$ and shows that irradiation has improved $J_c$ by a factor 3 or 4.

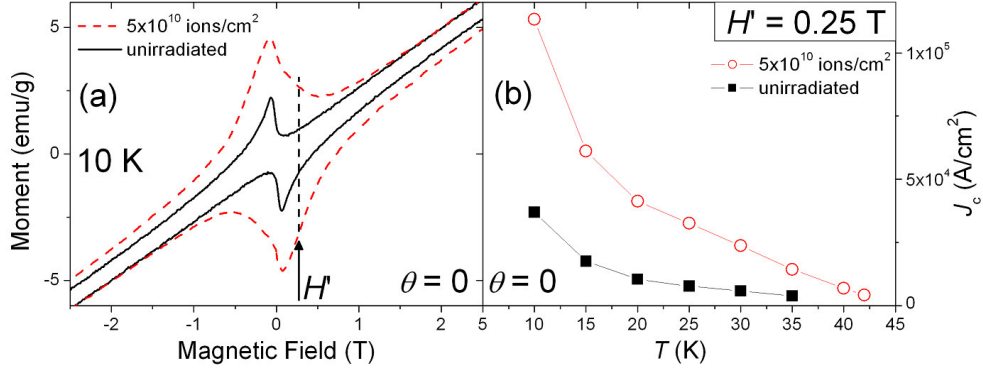

Figure 2. (a) Magnetization versus magnetic field (*M-H*) loops at 10 K on the unirradiated (black curve) and $5\times10^{10}$ ions/cm$^2$ irradiated fragments (red dashed curve). The field is applied parallel to the irradiation direction – defined as $\theta = 0$ deg. Pinning enhancement is most notable at fields below 1 T corresponding to the matching field. (b) The critical current density at $H=H'=0.25$ T versus temperature for the irradiated (open circles) and unirradiated sets (closed squares) for field alignment with the defect direction.

In Fig. 3(a) we show *M-H* loops at 10 K on the irradiated fragments as the field is tilted from parallel ($\theta=0$ deg) to perpendicular ($\theta=90$ deg) to the columnar defects. The different *M-H* loops show a clear systematic change in magnetization with $\theta$ for fields below the matching field. At fields close to zero the curves for different angles all intersect – i.e. the remnant magnetizations are similar. This effect has been attributed to 'flux flop' in the HTSC (16;17) where isotropic pinning was observed at low applied fields. The variation in $\Delta M$ with $\theta$ is greatest close to $H'=0.25$ T, and in Figs. 3(b) and 3(c) we show this variation in $\Delta M(\theta)$ at 10 K and 35 K, respectively. As expected the $\Delta M$ is a maximum for the parallel field alignment. The black squares in Figs. 3(b) and 3(c) show $\Delta M(\theta)$ for the unirradiated fragments, which confirm that there is no angular dependence in the unirradiated sample. Comparison of the $\Delta M(\theta)$ curves at 10 K and 35 K shows that at 10 K there is still a significant increase in pinning with irradiation for the field perpendicular, $H\perp$. At 35 K the pinning for the $H\perp$ case shows little enhancement with irradiation.

It is important to compare the effect that columnar defects have on pinning in the 1111 and 122 systems with similar studies that were originally done on the HTSC superconductors such as YBCO and BSCCO (18-20). Irradiation by heavy ions is known to have a maximum effect, as the produced columnar defects can pin the line-like vortices along their entire length. This leads to a directional enhancement of $J_c$ which is dependent on the angle between the applied field and the direction of defects. In particular a lock-in effect is expected for small field angles away from the defect direction and at high temperatures (21), similar to the lock-in effect of pinning by twin planes in YBCO (22). This directional enhancement was observed down to ~20 K in YBCO crystals which have a moderate superconducting anisotropy $\gamma$, but as $\gamma$ increases, (such as for highly anisotropic BSCCO crystals) there evolves a temperature crossover to a regime where the pinning enhancement is isotropic (16). This effect was attributed either to the increasing dominance at low temperature of pinning by random (therefore isotropic) point defects, or to a 3-D to 2-D crossover in the vortex structure with decreasing temperature. For BSCCO the pancake vortices interact to form a line-like structure at high temperature but decouple at low temperature, and therefore lose any preference for directional pinning.

Measurements of $J_c$ and magnetic relaxation as a function of temperature and angle have in the past (23;24) provided insight into the type of pinning mechanism that is responsible and addressed whether the mechanism is significantly changed by the introduction of columnar defects. Figure 3 implies that the pinning enhancement with columnar defects becomes more isotropic at low temperatures reminiscent of pinning in YBCO and BSCCO, except that no complete crossover regime is observed for the 1111 material. In this regard the 1111 oxypnictides appear to be closer in behaviour to YBCO.

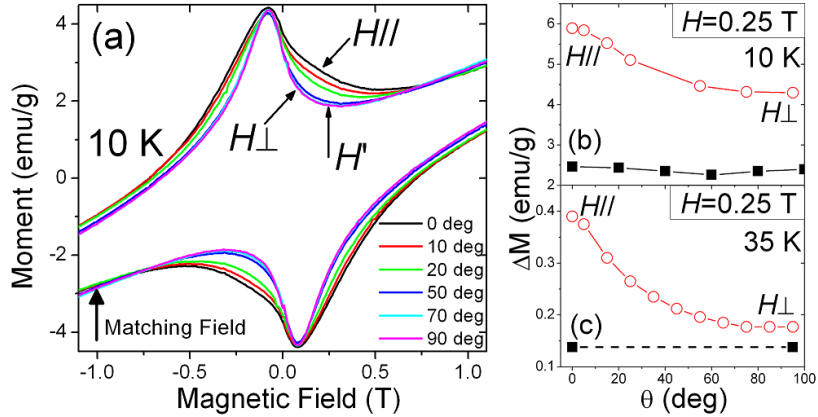

Figure 3. (a) *M-H* loops of the $5\times10^{10}$ irradiated fragments as the applied field is tilted from parallel ($\theta=0$ deg) to perpendicular (90 deg) to the irradiation direction at 10 K. $\Delta M$ is extracted at the field *H'* indicated in (a) and plotted versus $\theta$ in (b) at 10 K and (c) at 35 K for $5\times10^{10}$ irradiated fragments (open circles) and the unirradiated fragments (closed squares).

Figure 4 shows the field dependence of the normalised creep rate at the temperatures 20 K and 35 K. The magnetization signal from the set of unirradiated fragments was too small to obtain reliable creep data, however the magnetisation behaviour of the fragments was characteristic of that of the bulk, bar sample (6). Therefore, the creep data was obtained on the bar sample where a more reliable signal could be taken. (The large error bars for the irradiated curve shown in Fig. 4 reflect that the set of fragments was much smaller in mass than the unirradiated bar sample.) Figure 4 reveals that the creep rate in the irradiated fragments at 35 K is suppressed at fields below the matching field compared to the unirradiated bar. This is consistent with previous studies where the columnar defects suppressed vortex creep as the defects were effective pinning sites. At 20K the creep rate in the irradiated and unirradiated cases is similarly low (within the error bars), increasing slowly with field from the low-field value ~0.03, confirming that the columns are having much less direct influence on the pinning at low temperatures.

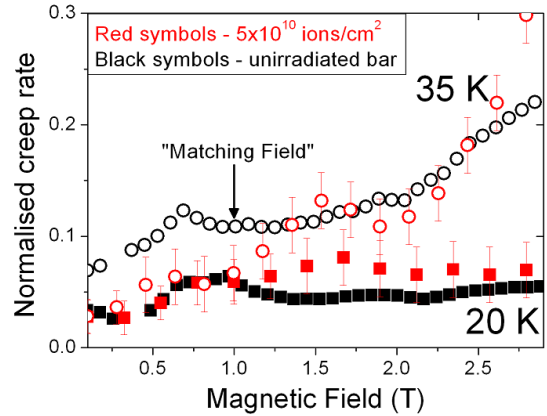

Figure 4. Normalised creep rate, *S*, versus applied field at 20 K (closed symbols) and 35 K (open symbols) taken on the unirradiated bar and the $5\times10^{10}$ irradiated fragments. At 20 K the creep rate for the two samples is unchanged, however at 35 K the creep in the irradiated fragments below the matching field is suppressed compared to the creep in the unirradiated bar sample. The peak feature in *S* below 1 T for the unirradiated bar was associated with the peak effect in the *M-H* loop (6).

Previous studies of magnetic relaxation (3;25) show that the relaxation rates are large in the 1111 phase and suggest a collective vortex pinning scenario. The vortex *H-T* phase diagram has been mapped and this shows a large region where the melting of the vortex lattice occurs for both 1111 and 122 families (10;26). The vortex behaviour is described as being intermediate of YBCO and BSCCO, but perhaps closer to YBCO. However, this point is still controversial as it has also been reported (27) from measurements of the flux dynamics in the 1111 system that a 2-D vortex behaviour, comparable to the pancake vortex structure in BSCCO 2223 superconductor, was observed. Similar to the Nakajima *et al.* work (4), we find a strong suppression in vortex creep rate below the matching field at high temperatures, which is consistent with the description of the vortex phase diagram obtained from unirradiated samples.

It is of some surprise that the columnar defects act as less effective pinning sites in the 1111 system than for the recent report of pinning enhancement in 122 due to columnar defects. As the 1111 is considered to be more anisotropic than 122, we would have expected greater enhancement due to the columnar defects. Complication in direct comparison is introduced however because the 1111 system we have studied is polycrystalline and randomly oriented and although we do not believe that the connectivity in the fragments is impaired by irradiation, the random orientation is likely to be the dominant cause of the reduced pinning.

## 4. Conclusions

We have studied the effect of heavy ion irradiation with 2 GeV Ta ions in polycrystalline $NdFeAsO_{0.85}$ and its effect on the critical current density $J_c$ and the flux dynamics. We showed the formation of continuous collinear columnar defects by using high resolution TEM. The irradiation has marginally (factor of 3 to 4) increased the $J_c$ to $\sim 10^5$ A/cm$^2$ at 10 K. By comparison, heavy ion irradiation in the 122 system, $Ba(Fe_{0.93}Co_{0.07})_2As_2$, saw $J_c$ increased by a factor ten to $\sim 10^6$ A/cm$^2$ at the same temperature, still a small increase compared to the observations in HTSC. We observed a directional enhancement to the pinning by columnar defects, with a peak in $J_c$ for $H$ parallel to the defects. The directional pinning enhancement becomes more isotropic at low temperatures, but without an actual crossover to isotropic pinning, which is similar to the effect of columnar defects in YBCO. The stronger pinning by heavy ion tracks observed at the higher temperatures is accompanied by a suppression of the creep rates for fields below the matching field of 1 T. Similar improvement in flux pinning was observed with neutron irradiation in polycrystalline $SmFeAsO_{1-x}F_x$ where the defects are point-like. Our work suggests irradiation studies in single crystals of 1111 oxypnictides would certainly be beneficial. The small pinning enhancement so far reported in oypnictides due to columnar defects compared to cuprate system requires further detailed exploration.

**Acknowledgements**

The work at Imperial College London is supported by the Engineering Physical Science Research Council, Grant EP/E016243/1. Y.Y. also acknowledges partial support by The Israel Science Foundation (ISF), Grant 499/07.